# Title:

# Reprogrammable polaritonic skyrmion and meron lattices based on the phase-change material $In_3SbTe_2$

Author(s) and Corresponding Author(s)*:

Lina Jäckering*,+,[1,2], Arvid Schick+,[1,2], Enrico Baù+,[3], Lukas Conrads[1,2], Kenji Watanabe[4], Takashi Taniguchi[5], Christoph Stampfer[2,6,7], Matthias Wuttig[1,2,8], Andreas Tittl[9,3], Thomas Taubner*,[1,2]

jaeckering@physik.rwth-aachen.de; taubner@physik.rwth-aachen.de

+authors contributed equally

## Affiliations

[1]1st Institute of Physics (IA), RWTH Aachen University, 52074 Aachen, Germany

[2]Juelich-Aachen Research Alliance (JARA-FIT), 52425 Juelich, Germany

[3]Nanoinstitute Munich, Faculty of Physics, Ludwig-Maximilians-Universität München, 80539 München, Germany

[4]Research Center for Electronic and Optical Materials, National Institute for Materials Science, 1-1 Namiki, Tsukuba 305-0044, Japan

[5]International Center for Materials Nanoarchitectonics, National Institute for Materials Science, 1-1 Namiki, Tsukuba 305-0044, Japan

[6]2nd Institute of Physics, RWTH Aachen University, 52074 Aachen, Germany

[7]Peter Grünberg Institut (PGI-9), Forschungszentrum Jülich, 52425 Jülich, German

[8]Peter Grünberg Institut (PGI-10), Forschungszentrum Jülich, 52425 Jülich, Germany

[9]Institute of Photonics, Hamburg University of Technology, 21073 Hamburg, Germany

## Abstract

Optical topological quasiparticles such as skyrmions and merons consist of engineered electromagnetic fields which can be realized via the controlled interference of polaritons. The resulting polaritonic topological lattices offer a promising platform for on-chip information encoding and nanoscale light-matter interaction. However, realizing these functionalities requires active tailoring of the polaritonic topological lattice. Although previous works have achieved tuning of the size of the polaritonic lattice sites, these methods relied on cumbersome and time-consuming fabrication processes and do not allow reconfiguration of the topological invariant. Here, we demonstrate fast optical prototyping of reconfigurable hexagonal and square coupling structures for the generation of polaritonic topological lattices with hyperbolic phonon polaritons in hBN using the phase-change material $In_3SbTe_2$. As $In_3SbTe_2$ can be switched reversibly between its dielectric ($\varepsilon > 0$) amorphous and metallic ($\varepsilon < 0$) crystalline phase by laser pulses, metallic $In_3SbTe_2$ coupling structures can be optically programmed and reconfigured. Our approach enables tailoring the size of the topological lattice sites by tuning the incident frequency. Moreover, $In_3SbTe_2$ allows for tailoring the topological invariant by reprogramming the coupling structure. Our platform enables reprogramming of robust optical topologies within minutes, opening a route toward rewritable on-chip photonic computing based on robust and on demand reconfigurable topological states.

## Introduction

Skyrmions are quasiparticles with stable topological textures that originate from topological theory of elementary particles developed in 1961 by Skyrme[1]. A skyrmion is a three-dimensional vector field confined in a two-dimensional space, defining a mapping from real space to the unit sphere in parameter space. The skyrmion number is the topological invariant and refers to the number of times the vector field wraps around a unit sphere.[1,2] The skyrmion number classifies the resulting topology: a quasiparticle with skyrmion number $S_T = 1$ is termed a skyrmion and with skyrmion number $S_T = 0.5$ a meron. In particular, skyrmions appear in a wide range of physical systems[1] ranging from platforms in condensed matter physics, such as Bose-Einstein condensates[3], magnetism[4–7] and twistronics[8] to more recent realizations in optics[2,9].

Optical topological quasiparticles are promising for applications in ultrafast vector imaging[10], nanoscale metrology[11], deep-subwavelength microscopy[9], topological Hall bar devices[12], and optical information technology[13–15]. One way to realize optical topological quasiparticles is through controlled interference of polaritons. Polaritons are quasiparticles arising from the hybrid excitation of photons and collective dipole excitations in a material, e.g. plasmons in a metal or phonons in a polar crystal. Initially, polaritonic skyrmions have been realized with surface plasmon polaritons (SPPs) at a gold-air interface[2,9,16]. To realize the controlled interference with controlled phase, coupling structures with sophisticatedly designed boundaries to launch the polaritons are utilized. For example, Tsesses et al.[2] illuminated a hexagonal coupling structure consisting of 6 gratings arranged in a hexagon to launch the SPPs with circularly polarized light. For controlled interference, one of the gratings was offset by half of the SPP wavelength to compensate for the phase shift induced by the circularly polarized light. However, these SPPs generally feature a dispersion close to that of free-space light and hence a low spatial confinement. Recently, phonon polaritons with much higher confinement and lower losses have been exploited for the generation of polaritonic topological lattices[17–20]. Further, they enable tuning the size of the topological lattice site through small changes of the incident frequency (<10%) [17,18]. However, these approaches require a different coupling structure for each frequency, as the design of these structures depends on the operation frequency [16–18]. Thus, for every operating frequency a new structure has to be fabricated, limiting the applicability of polaritonic skyrmions for tunable on-chip computing. To enable same-structure tunability of the size of individual skyrmions, Baù et al.[21] employed a metasurface supporting quasi-bound states in the continuum resonances with homogeneous fields across the single resonator structures that generate the polaritonic topological lattices. By tuning the frequency within the quasi-bound states in the continuum resonance, tuning the size of the topological lattice site has been demonstrated. However, these methods suffer from a shared reliance on cumbersome and time-consuming fabrication schemes. Moreover, they do not enable reversible switching between different topological quasiparticles, such as polaritonic skyrmion and meron lattices. Such active tuning of the topological charge is essential for future on-chip information processing devices and requires modifications in the launching structure of the lattice.

A promising platform for active tuning of topological charges and fast prototyping of coupling structures are phase-change material (PCMs). PCMs can be reversibly switched between two (meta)stable phases, the amorphous and crystalline phase[22,23] with distinct optical properties. Among PCMs that usually feature a dielectric amorphous and crystalline phase in the infrared spectral range, the PCM $In_3SbTe_2$ (IST) stands out as it can be switched between a dielectric ($\varepsilon > 0$) amorphous and a metallic ($\varepsilon < 0$) crystalline phase[24]. As metallic ($\varepsilon < 0$) polaritonic launching structures can directly be imprinted by local crystallization, IST has been introduced as a versatile platform for polariton optics[25–28]: Polariton optics programmed into IST have been demonstrated for surface phonon polaritons in SiC[25], surface plasmon polaritons in CdO[26], out-of-plane hyperbolic phonon polaritons in hBN[27,29], and in-plane hyperbolic phonon polaritons in $\alpha$-$MoO_3$[29,30].

Here, we utilize IST to achieve fast prototyping of reconfigurable coupling structures allowing a) for reconfiguration of the coupling structure and hence of the topological charge, b) for coupling over a broad spectral range to enable lattice site size tuning without reconfiguration, and c) for coupling with linearly polarized light and detection in reflection mode, bypassing the need for additional optical elements for the generation of circularly polarized light such as waveplates. Optically we program hexagonal and square coupling structures into the PCM IST through the hBN flake supporting hyperbolic polaritons to generate polaritonic skyrmion and meron lattices. Reconfiguring the shape and size of the coupling structure allows for tuning the topological charge and size of individual lattice sites, characterized with near-field optical microscopy. By tuning the wavenumber, we decrease the lattice site size and achieve a maximum spatial confinement of λ/40.

## Results

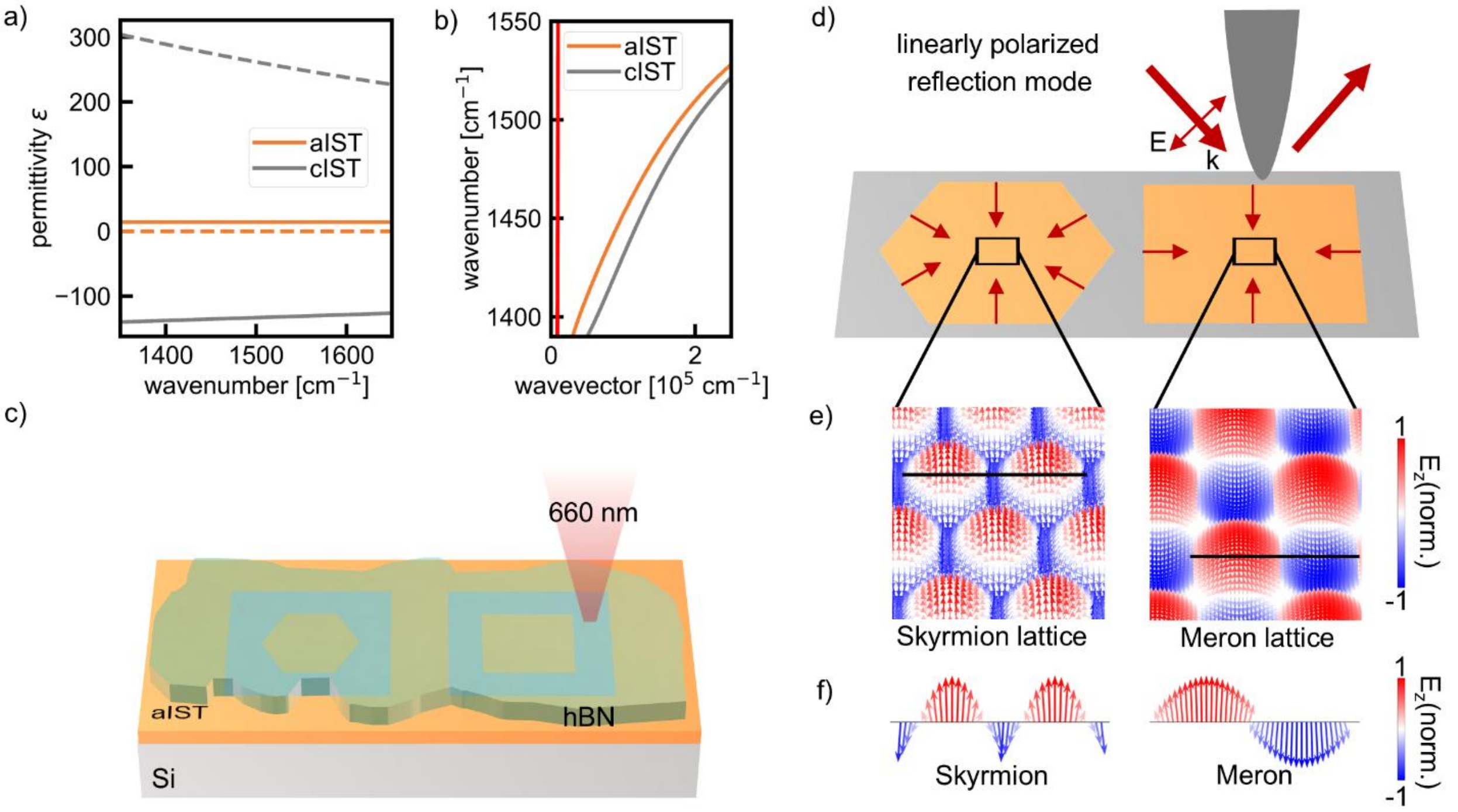

***Figure 1: Direct optical programming of coupling structures for phonon polaritonic topological lattices into IST.*** *a) Real (solid) and imaginary (dashed) part of the permittivity of amorphous (orange) and crystalline (grey) IST. b) Dispersion of hBN's phonon polaritons with amorphous (aIST; orange) and crystalline (cIST; gray) as a substrate together with the dispersion of light (red). c) Sketch of the optical programming with a visible laser (red) of hexagonal and square crystalline IST (grey) coupling structures through the hBN flake (green) into amorphous IST (orange). d) Schematic of characterizing phonon polaritonic topological lattices with s-SNOM in reflection mode and illumination with p-polarized light. e) Schematic of vectorial* $E_z$ *field of the phonon polaritonic topological lattices arising in the different coupling structures sketched in d),*

*yielding a skyrmion lattice for hexagons and a meron lattice for square coupling structures. f) Line cut through the polaritonic lattices in e) to visualize the type of topology.*

To achieve reprogrammable optical topologies, we program coupling structures into the PCM IST to generate polaritonic topological lattices with hyperbolic phonon polaritons in hBN (see Supplementary Note 1) that we investigate with scattering-type scanning near-field optical microscopy (s-SNOM). In the frequency range investigated, the two phases of IST feature a vast permittivity contrast. While IST in its dielectric amorphous phase (aIST) has a positive permittivity, in its metallic crystalline phase (cIST) it has a strongly negative permittivity ($\varepsilon < 0$) owing to its Drude like behavior (c.f. Figure 1a). Hence, polariton launching structures to realize coupling structures can be programmed by locally crystallizing the IST[25–27,29,30]. The phonon polariton dispersion of hBN is strongly influenced by the substrate, leading to distinct dispersion behavior for hBN supported on aIST and cIST, as depicted in Figure 1b and demonstrated experimentally previously[27,29]. Importantly, hyperbolic polaritons are highly confined and the polariton wavelength varies strongly upon the incident frequency[31,32] allowing for tuning the size of the topological lattice sites similarly to SiC membranes[17] and for even higher confinement ($\lambda_0$/50- $\lambda_0$/100)[31] compared to SiC (~$\lambda_0$/4)[17].

We exploit laser-induced crystallization to optically program coupling structures of various shapes into the IST through hBN flakes exfoliated on top of IST (c.f. Figure 1c). Our fabrication scheme consists of only three steps: sputtering of the amorphous IST thin film, the hBN flake exfoliation onto the IST, and the optical programming of coupling structures by local laser pulses for crystallization of the IST. Thereby, the layer stack is realized within a few hours and the polariton coupling structures are optically programmed within minutes[30]. Further coupling structures can be added to the very same flake or to flakes on the same sample within minutes and the structures can be reconfigured within seconds[30]. Remarkably, our approach enables on demand reconfigurability of previously programmed coupling structures. Exemplarily, we sketch two different shapes of coupling structures in Figure 1c.

To investigate the polaritonic topological lattices generated with these coupling structures, we utilize s-SNOM [33–35] (see Methods). In s-SNOM a sharp tip is brought in proximity to the sample surface and illuminated with infrared light. Upon illumination the tip acts as an optical antenna and can therefore enhance and scatter local near fields, such as polaritons [32,35–39]. Importantly, we use linearly polarized light and employ s-SNOM in reflection mode as sketched in Figure 1d. Thereby, our detection scheme contrasts the previously used combination of circularly polarized light and s-SNOM in transmission mode[2,16,17]. With our detection scheme we overcome the need for additional frequency dependent optics such as waveplates or polarizers to generate circularly polarized light. As our coupling structures with side length of a ≈ 1-3 μm are smaller than the incident wavelength ($\lambda_0 \approx$ 6.7-6.8 μm), we observe similar launching amplitudes from all edges and similar phases from some edges despite the oblique illumination with linearly polarized light from top. Thus, not only our fabrication but also our detection scheme stands out due to its simplicity and faster turn-around times compared to previous approaches.

The two coupling structures shown in Figure 1d generate a different kind of topological lattice[18]. As visualized by the schematic electric field vectors $E_z$ in Figure 1e, a hexagonal coupling structure generates a skyrmion lattice, and a square coupling structure a meron lattice. The line cuts (c.f. Figure 1f) through the vectorial electric field of the topological lattices in Figure 1e highlight the different topological charges. In the hexagonal skyrmion lattice, the normalized electric field $E_z$ change from -1 to 1 within one lattice site, thus wrapping the entire unit sphere and hence the skyrmion number is $S_T = 1$[1]. For the square meron lattice, the skyrmion number is $S_T = 0.5$, corresponding to a meron[1].

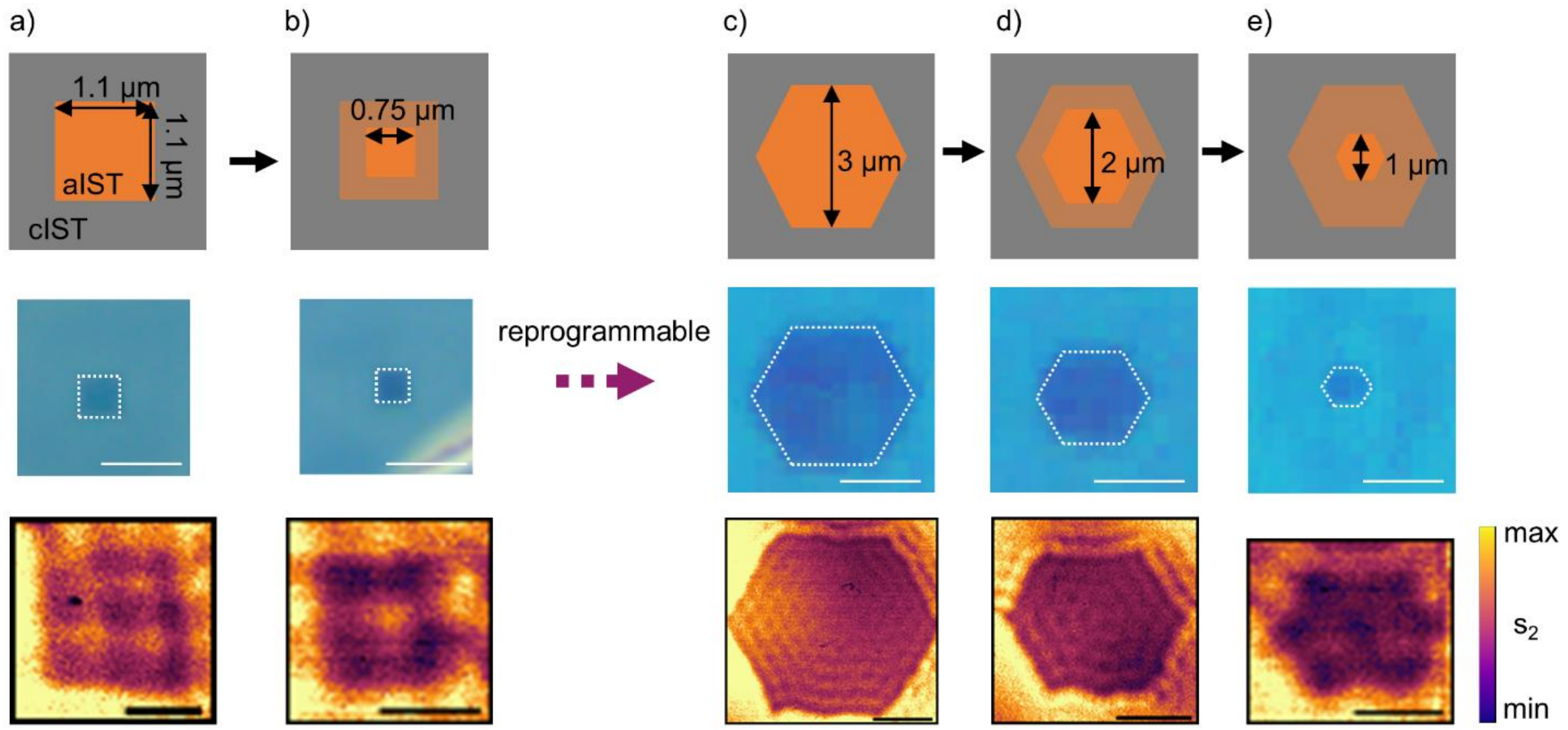

***Figure 2: Reconfiguring shape and size of coupling structures to tailor topological charge and lattice.*** *Sketch (top), optical light microscope image (middle) and near-field optical amplitude image(bottom) of a) and b) square coupling structures to tailor the polaritonic topological lattice at 1480 cm$^{-1}$ ($d_{hBN}$=61 nm) and c) - e) hexagonal coupling structure to tailor topological lattices at 1500 cm$^{-1}$($d_{hBN}$=48 nm). Scalebars in the optical light microscope image are 2 µm. Scalebars in the s-SNOM amplitude images in a), b) and e) are 0.5 µm, in c) and d) are 1 µm.*

Our optical programming allows for on demand programming and reconfiguration of the coupling structure. In Figure 2 we show optically programmed coupling structures of different shapes and sizes. For each structure, we show a sketch of the fabricated structure at the top, a light microscope image in the middle and the measured s-SNOM amplitude image at the bottom. In Figure 2a top, we optically program a square coupling structure into the IST by precisely overlapping laser pulses (λ=660 nm) for crystallization and leaving an amorphous area of 1.1 µm x 1.1 µm as indicated in the sketch (see Methods for details). In the optical light microscope image in Figure 2a middle, the amorphous square can be identified as it appears dark in the bright crystalline surrounding. The s-SNOM amplitude image (c.f. Figure 2a bottom) reveals four spots of high amplitude in the corners of the square structure that we attribute to the interference of the phonon polaritons in hBN launched at the edges of the IST square structure. Due to the small size of the structure, the phonon polaritons in hBN are confined in the region of amorphous IST. Hence, the same structure can also be interpreted as a square resonator for phonon polaritons in hBN similar to previously demonstrated optically programmed circular resonators in IST for phonon polaritons in hBN[27].

Demonstrating the reconfigurability, the side length of the square is decreased by further crystallization (c.f. sketch of the smaller square in Figure 2b top). This smaller square of about 0.75 µm x 0.75 µm also appears dark in the optical light microscope image in Figure 2b middle. Remarkably, the local polaritonic field pattern in the s-SNOM amplitude image (c.f. Figure 2b bottom) taken at the same illumination wavenumber is different to that of the larger square structure: In the center of the amorphous square, we observe a single spot of high amplitude. Decreasing the size of the square structure changes the interference patterns of the phonon polaritons. By reamorphization and additional crystallization, in principle, the complete shape of the structure could be reprogrammed.

In Figure 2c-2e we use a similar optical programming scheme but leave out a hexagonal amorphous area. By additional crystallization we decrease the side lengths of the hexagonal coupling structure subsequently from 3 µm to 1 µm as indicated in the sketches and visible in the optical light microscope images. We perform s-SNOM amplitude images of the differently sized hexagonal coupling structures at the same illumination wavenumber. These s-SNOM amplitude images reveal distinct polaritonic interference patterns. While for two larger hexagonal coupling structures multiple spots of high amplitude are observed (c.f. Figures 2c and 2d bottom), for the smallest hexagonal coupling structure there is only one spot of high amplitude in the center (c.f. Figure 2e bottom). Our s-SNOM amplitude images show that by crystallizing the material with additional laser pulses we can reprogram the polaritonic interference pattern without changing the illumination frequency by changing the coupling structure, i.e., the launching structures. To demonstrate the on demand switching between different topological charges that is enabled through our material platform, we evaluate the polaritonic interference pattern emerging from a square and hexagonal coupling structure in the picture of polaritonic topological lattices in the following.

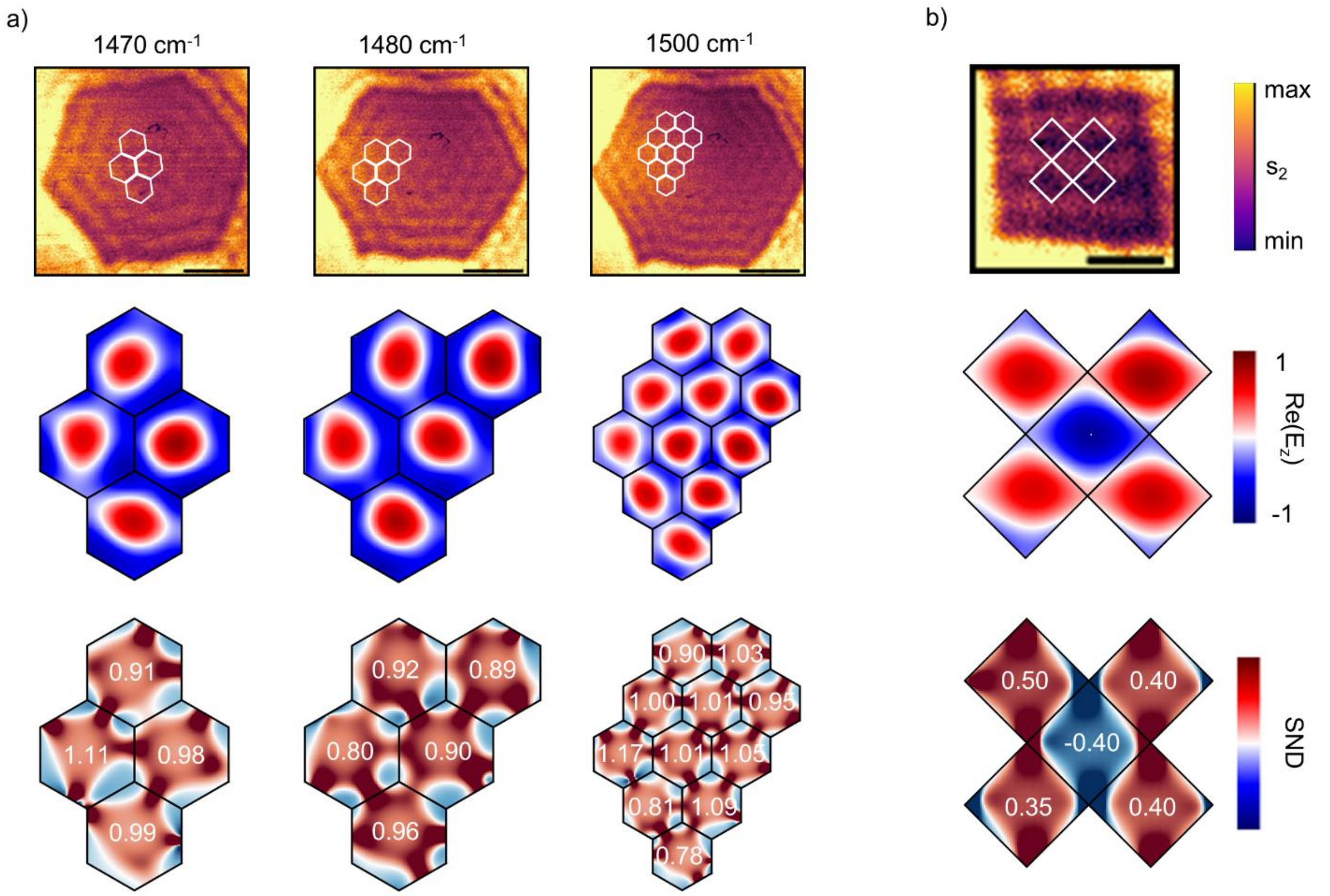

***Figure 3: Tunable polaritonic skyrmion and meron lattices.*** *Near-field optical amplitude image top), filtered electric field in z-direction* $E_z$ *(middle) and topological invariants* $S_T$ *(white numbers) and the color-coded skyrmion number density (SND) (bottom) of a) polaritonic skyrmion lattices at three different frequencies and b) a polaritonic meron lattice at 1515 cm*$^{-1}$*. Scalebars in a) correspond to 1 µm and in b) to 0.5 µm.*

To visualize the polaritonic interference pattern and determine the topological invariant, we employ s-SNOM for a hexagonal coupling structure at three different frequencies and for a square coupling structure (c.f. Figure 3a and 3b). In the polaritonic interference pattern, we mark selected lattice sites in white whose evaluation will be shown in the following. Owing to the polariton dispersion, the lattice site size decreases from 239 nm for the hexagonal structure at 1470 cm$^{-1}$ to 168 nm at 1500 cm$^{-1}$. Importantly, in s-SNOM we measure predominantly the near-fields in z-direction and acquire both amplitude and phase. This allows us to extract the complex electric field in z-direction $E_z$. For evaluation of the topology we follow the approach of ref.[17]: the extracted $E_z$ is Fourier-filtered to isolate the polaritonic contributions (see Supplementary Note 2). The filtered real part of the electric field Re($E_z$) of selected lattice sites (white frames in Figure 3 top row) in the polaritonic interference pattern is plotted in Figure 3a and 3b mid row, for the hexagonal and square lattice, respectively. We observe similar behavior in the lattice sites of the hexagonal lattice with positive Re($E_z$) in the center of the lattice sites and negative Re($E_z$) at the boundaries (c.f. Figure 3a mid row) whereas we observe lattice sites of almost constant field alternating in sign for neighboring lattice sites in the square lattice (c.f. Figure 3b mid row). From the Fourier filtered Re($E_z$), we see a continuous rotation of the electric field in z-direction within the lattice sites for the hexagonal and for neighboring lattice sites for the square lattice, which is key for the observation of polaritonic topological lattice.

Next, we calculate the skyrmion number $S_T$ (see ref.[17] and Supplementary Note 2 for details). From the filtered, complex electric field in z-direction $E_z$, the in-plane electric field is calculated using Maxwell's equations and boundary conditions (c.f. Supplementary Note 3). Then, the skyrmion number density is calculated via $SND = \frac{1}{4\pi}\hat{e}\left(\frac{\partial \hat{e}}{\partial x} \times \frac{\partial \hat{e}}{\partial y}\right)$ with $\hat{e} = \frac{E(r,t)}{|\mathrm{E}(r,t)|}$ and displayed for selected lattice sites in Figure 3 bottom. We integrate the SND across the lattice site and find a skyrmion number $S_T$ of about 1 for the hexagonal polaritonic lattice and a $S_T$ of about 0.5 for the square meron lattice, demonstrating that the optically programmed structures can generate skyrmion and meron lattices depending on the shape of the coupling structure. The quality of the polaritonic topological lattices is of similar quality as previously realized polaritonic topological lattices employing conventional fabrication schemes which is for most skyrmion lattice sites roughly within 10% away from the ideal values of $\pm 1$[21]. Further, we demonstrate robustness of the skyrmion lattice against global phase perturbations in Supplementary Note 4. In Figure 3a, the hexagonal skyrmion lattice of the very same coupling structure shows differently sized lattice sites for the three different incident wavenumbers. Remarkably, the lattice site size is tailored without adapting the coupling structure as our coupling structures are independent of the polariton wavelength, contrasting previous approaches in literature[17].

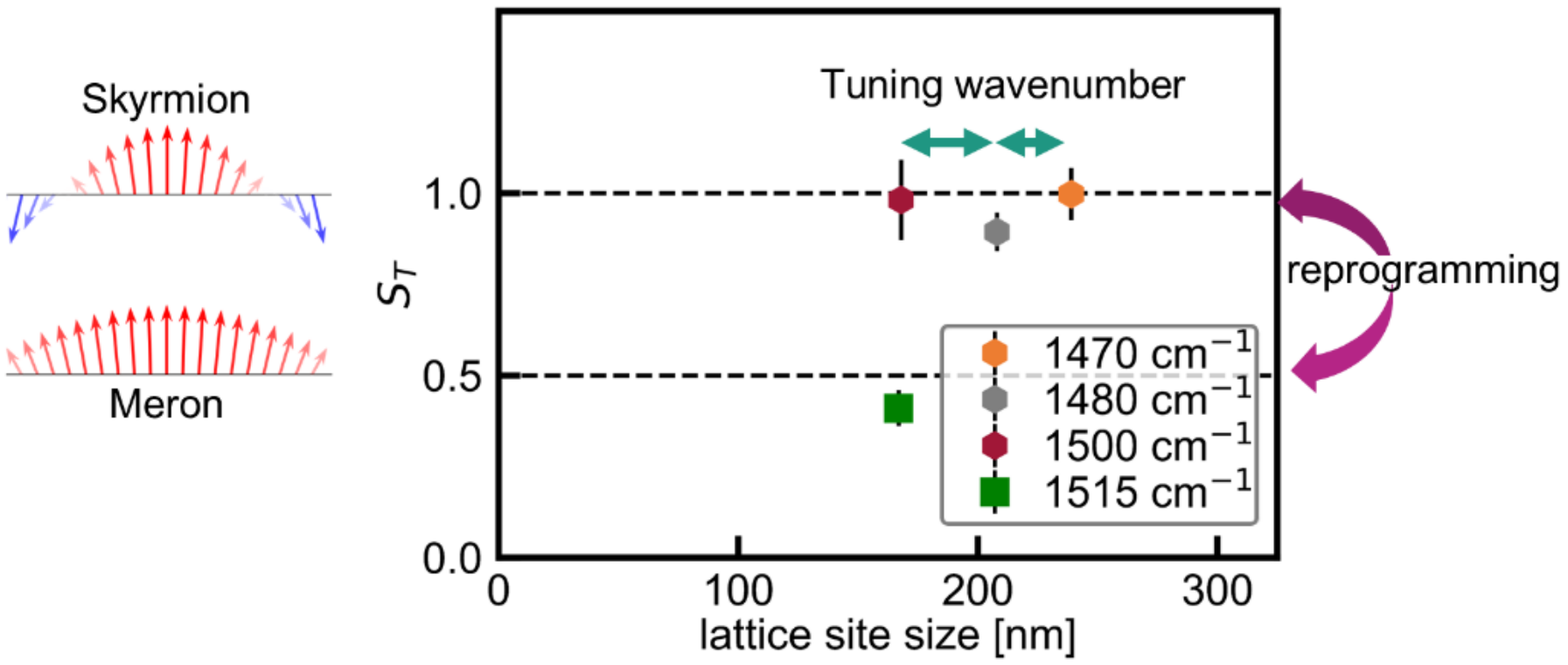

***Figure 4: Tailoring topology and lattice site size of polaritonic topological lattice.*** Experimentally extracted absolute value of skyrmion number $S_T$ (c.f. Figure 3) in dependence on the lattice site size. $S_T$ extracted from the hexagonal coupling structure and from the square coupling structure are displayed as hexagons and squares, respectively. Our platform allows for two distinct tailoring mechanisms: tailoring the lattice site size by tuning the incident wavenumber without reconfiguration of the coupling structure. Tailoring of the topological invariant by reprogramming the coupling structure.

Our fast prototyping of coupling structures relying on the PCM IST stands out as it allows for two simple tailoring mechanisms at the same time as we highlight in Figure 4. We plot the mean value of the absolute value of the skyrmion number $|S_T|$ of the lattice sites evaluated in Figure 3 against the lattice site size. The error bars represent the respective standard deviation. First, by tuning the incident frequency, we can tune the size of the topological lattice (c.f. blue arrow in Figure 4), importantly, without adaption of the coupling structure as we demonstrate for a hexagonal coupling structure in Figure 3. By increasing the incident frequency, we decrease the lattice site size of the hexagonal lattice to 168 nm. This corresponds to a confinement of $\lambda_0/40$ of the skyrmion compared to the incident wavelength. Secondly, by reprogramming the coupling structure and changing its shape, in principle the topological invariant can be tailored. When reprogramming a hexagonal coupling structure into a square coupling structure, the topological invariant can be changed from 1 to 0.5, i.e. the skyrmion is turned into a meron (c.f. magenta arrow in Figure 4). Therefore, reprogramming of the coupling structure allows for changing the topological charge in a polaritonic topological lattice.

## Discussion

In summary, we demonstrate a platform for fast prototyping of reconfigurable polaritonic topological lattices based on the PCM IST and hyperbolic phonon polaritons in hBN. We optically program coupling structures of various sizes and shapes for polaritonic skyrmion and meron lattices. We find that, in contrast to previous investigations of polaritonic topological lattices, s-SNOM can be used in reflection mode and with linearly polarized light to visualize polaritonic topological lattices. This is especially advantageous, as no -usually frequency dependent-waveplates are needed, enabling investigation over a broad spectral range. Our platform allows for tailoring the size of the topological lattice site by changing the incident frequency without the need to adapt the coupling structure, in stark contrast to previous demonstrations[2,9,16,17]. Further, our platform allows for tailoring the topological invariant through on demand reprogramming of the coupling structures, e.g., turning a hexagon into a square and hence on demand tuning from a skyrmion lattice to a meron lattice. Our approach can be straightforwardly expanded to include a broad range of more complex topologies, such as skyrmion bags[40–42] or topological charge vectors[43]. Our platform speeds up fabrication and facilitates investigation of polaritonic topological lattices and hence brings advances to on-chip mid-IR information encoding and sensing based on topological devices. We envision our platform brings prospects for active control of topological properties of light and on-demand control over the topological properties of light, capabilities that are key for the exploration of novel light matter interactions and promising for applications in spin-optics, imaging and quantum technology[1].

## Methods

**Sample Fabrication.** Using direct current and radio frequency sputter deposition, a 50 nm thick layer of amorphous $In_3SbTe_2$ and a 20 nm thick capping layer of ZnS:$SiO_2$ were applied on a Si substrate. The capping layer (80% ZnS, 20% $SiO_2$) prevents the sample from oxidation. Flakes of hBN were then directly exfoliated onto the previously described amorphous PCM sample. The investigated hBN flakes are 48 (hexagons) and 61 nm (square) thick.

**Optical switching.** A home-built laser switching setup was used to crystallize the PCM locally through the already deposited hBN flake. Here, the light of a 660 nm laser diode is focused on the sample with a 10-fold objective with NA = 0.25. Each individual crystallized spot for the resonator structures was created by applying 100 pulses with a power of 36 mW and a pulse length of 500 ns. Due to the elliptical beam profiles the crystallized spots are elliptical. One elliptical crystallized spot has a lateral extension of approximately 1 x 1.5 $\mu m^2$. The sample was placed on a Thorlabs piezo stage (Thorlabs MAX311D/M with BPC303 piezo controller) allowing for precise movements of the sample. The amorphous resonator is created by placing multiple crystalline spots next to each other and leaving out a amorphous region in the center.

**s-SNOM.** In s-SNOM, a sharp metallic tip in proximity to the sample surface is illuminated by a laser. The tip, acting as an optical antenna, strongly enhances near-fields at the tip's apex. The scattered light scattered by the tip is detected interferometrically to extract amplitude and phase. s-SNOM is sensitive to local changes in the permittivity of the sample and to local near-fields[35]. It can directly probe the near-fields of the polaritons[32,36–39]. We used a commercially available s-SNOM (NeaSNOM neaspec GmbH) to image the polaritons in hBN on IST. We combined commercially available quantum cascade lasers (MIRcat-QT Mid-IR Laser) with a liquid nitrogen-cooled Mercury Cadmium Tellurium detector to investigate polaritons in the wavenumber range from 1470 to 1515 $cm^{-1}$. We used the pseudoheterodyne detection scheme to obtain s-SNOM amplitude and phase. We performed our measurements using tapping amplitudes between 65 and 75 nm and demodulated the signal at the second demodulation order of the tip-oscillation frequency.

## Code Availability

The computer codes developed for this study are available from the corresponding author upon request.

## Supporting Information

Supplementary Notes (PDF) on phonon polaritons in hBN, on the workflow for extracting the topological invariant from experimentally measured near-field data, on the relationship between the out-of-plane electric field and the corresponding in-plane field distribution, on the stability of experimentally measured skyrmion under global phase evolution, and on a comparative study of s-SNOM for characterization of polaritonic field topological lattices.

## Acknowledgements

The authors thank Asli Turan for the sputter deposition of the thin film layer stack and Hristiyana Kyoseva for exfoliation of the hBN flakes. L.J. acknowledges the support of RWTH University through the RWTH Graduate Support scholarship. L.J., L.C. and T.T. acknowledge support by the Deutsche Forschungsgemeinschaft (DFG No. 518913417 & SFB 917 “Nanoswitches”). M.W. acknowledges support by the Deutsche Forschungsgemeinschaft (SFB 917 “Nanoswitches”). E.B. and A.T. acknowledge funding from the European Union through ERC METANEXT (grant no. 101078018) and EIC NEHO (grant no. 101046329), from the Deutsche Forschungsgemeinschaft (DFG, German Research Foundation) under Germany’s Excellence Strategy (EXC 2089/2 –

390776260 and EXC 3120/1 – 533771286) and through the Emmy Noether Programme (TI 1063/1), and from the Center for NanoScience (CeNS) at LMU Munich. The views and opinions expressed are those of the author(s) only and do not necessarily reflect those of the European Union, the European Research Council Executive Agency or the European Innovation Council and SMEs Executive Agency. Neither the European Union nor the granting authority can be held responsible for them. K.W. and Ta.Ta. acknowledge support from the JSPS KAKENHI(GrantNumbers21H05233 and23H02052), the CREST (JPMJCR24A5), JST and World Premier Intern tional Research Center Initiative (WPI), MEXT, Japan.

**Author Contributions Statement**

L.J., L.C., E.B. and T.T. conceived the project with assistance by A.T.. A.S. performed the s-SNOM measurement. L.J. supervised the s-SNOM measurements. L.J., A.S., E.B. and L.C. analyzed the data. E.B. provided theoretical assistance and support. A.T. supervised E.B.. K.W. and Ta.Ta. provided high-quality hBN crystals. C.S. provided exfoliation equipment and van der Waals materials expertise. M.W. provided sputtering equipment and phase-change material expertise. All authors contributed to writing the manuscript.

**Competing Interests Statement**

The authors declare no competing financial interest.

# Supplementary Information:

# Reprogrammable polaritonic skyrmion and meron lattices based on the phase-change material $In_3SbTe_2$

Author(s) and Corresponding Author(s)*:

Lina Jäckering*,+,[1,2], Arvid Schick+,[1,2], Enrico Baù+,[3], Lukas Conrads[1,2], Kenji Watanabe[4], Takashi Taniguchi[5], Christoph Stampfer[2,6,7], Matthias Wuttig[1,2,8], Andreas Tittl[9,3], Thomas Taubner*,[1,2]

jaeckering@physik.rwth-aachen.de; taubner@physik.rwth-aachen.de

+authors contributed equally

## Affiliations

[1]1st Institute of Physics (IA), RWTH Aachen University, 52074 Aachen, Germany

[2]Juelich-Aachen Research Alliance (JARA-FIT), 52425 Juelich, Germany

[3]Nanoinstitute Munich, Faculty of Physics, Ludwig-Maximilians-Universität München, 80539 München, Germany

[4]Research Center for Electronic and Optical Materials, National Institute for Materials Science, 1-1 Namiki, Tsukuba 305-0044, Japan

[5]International Center for Materials Nanoarchitectonics, National Institute for Materials Science, 1-1 Namiki, Tsukuba 305-0044, Japan

[6]2nd Institute of Physics, RWTH Aachen University, 52074 Aachen, Germany

[7]Peter Grünberg Institut (PGI-9), Forschungszentrum Jülich, 52425 Jülich, German

[8]Peter Grünberg Institut (PGI-10), Forschungszentrum Jülich, 52425 Jülich, Germany

[9]Institute of Photonics, Hamburg University of Technology, 21073 Hamburg, Germany

**This PDF file includes:**

### Supplementary Note 1: Phonon polaritons in hBN

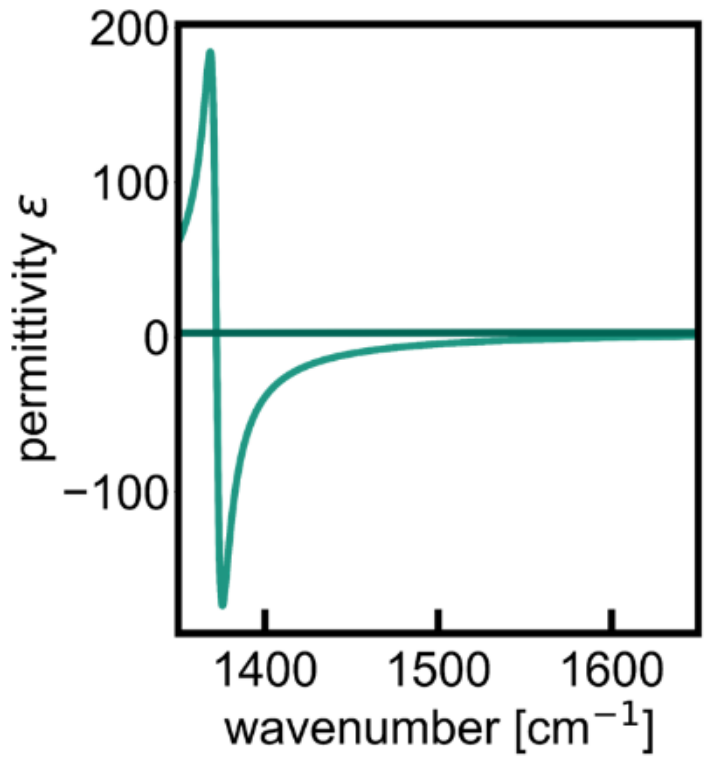

***Figure S2: Permittivity of hBN.*** *Real part of the in-plane (dark blue) and out-of-plane (light blue) permittivity of hBN.*

Hexagonal boron nitride (hBN) is a dielectric van der Waals material with covalent bonds in-plane and van der Waals bonds out-of-plane[1]. This highly anisotropic crystal structure also leads to optical anisotropy and hBN shows distinct permittivities in- and out-of-plane (c.f. Figure S1) [1]. Further, the resonances of the phonons in in- and out-of-plane direction are spectrally separated and thus are the corresponding reststrahlenbands. Figure S1a displays the permittivity of hBN in the upper reststrahlenband where the in-plane permittivity is negative while the out-of-plane permittivity is positive. These opposite signs define hBN as a natural hyperbolic material and cause a hyperbolic isofrequency contour for the phonon polaritons that can be excited within the reststrahlenband[1]. Hence, the polaritons are termed hyperbolic phonon polaritons. Importantly, these hyperbolic polaritons are highly confined, have low losses, and the polariton wavelength varies strongly upon the incident frequency [2,3]. Therefore, they are highly suitable for polaritonic topological lattices.

## Supplementary Note 2: Workflow for extracting the topological invariant from experimentally measured near-field data

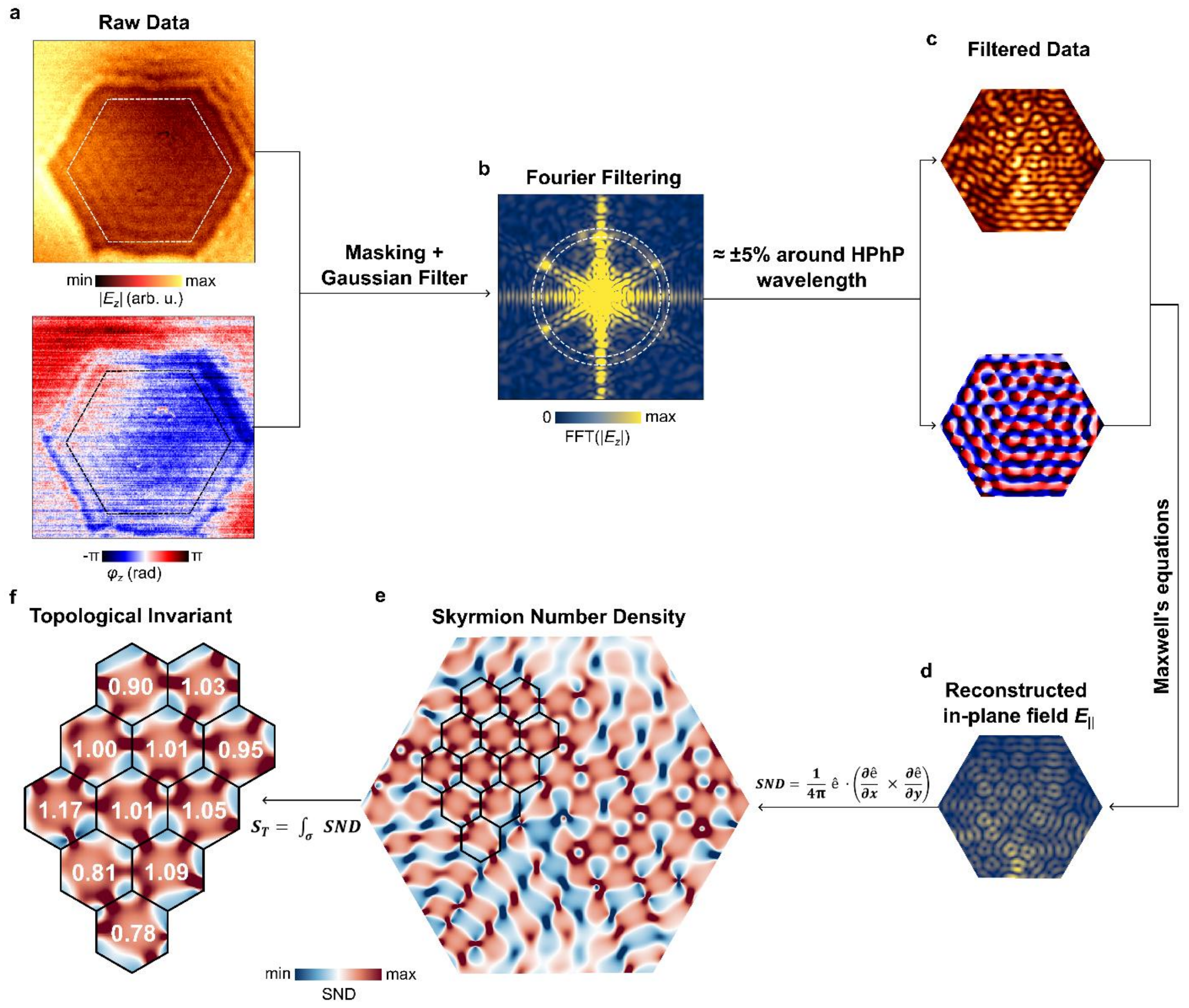

***Figure S2: Workflow for extracting the topological invariant from experimentally measured near-field data.*** *a) Raw s-SNOM measurements corresponding predominantly to the out-of-plane electric field amplitude* $|E_z|$ *(top, same as in Figure 3a right) and phase* $\varphi_z$ *(bottom). The dashed hexagon indicates the region of interest selected for subsequent analysis. b) Fourier transform of the measured complex field after applying a spatial mask and Gaussian apodization. A circular band-pass filter centered on the HPhP wavevector is applied, retaining only Fourier components within ±5% of the HPhP wavelength (equivalently, the corresponding in-plane wavevector). c) Inverse Fourier transformation yields the filtered out-of-plane field amplitude and phase, from which noise and unwanted spatial frequencies have been removed while preserving the polaritonic interference pattern. d) The in-plane electric field magnitude* $|E_{||}|$ *is reconstructed from the filtered out-of-plane field using the procedure described in Supplementary Note 3. e) The skyrmion number density SND is calculated from the normalized electric-field orientation ê, revealing the local topological charge distribution. The overlaid hexagonal grid defines the integration domains used for evaluating the topological invariant of individual unit cells. f) The skyrmion number* $S_T$ *is*

*obtained by integrating the skyrmion number density over each hexagonal unit cell with area σ. The resulting values, close to unity, confirm the presence of a robust skyrmion lattice.*

The analysis of the reprogrammable polaritonic topological lattices introduced in this work was performed using the out-of-plane complex electric field $E_z$. Amplitude and phase acquired in s-SNOM correspond predominantly to the out-of-plane complex electric field $E_z$ with amplitude $|E_z|$ and phase $\varphi_z$. To reliably extract the corresponding topological invariants, several processing and filtering steps were applied in both real and momentum space. First, both $|E_z|$ and $\varphi_z$ images obtained from s-SNOM (Figure S2a) are zero-padded to increase the sampling density in reciprocal space and thereby improve the apparent spectral resolution of the Fourier transform. Since the sharp edges introduced by thresholding and masking can lead to artificial high-frequency components due to the periodic assumption of the fast Fourier transform (FFT), a Gaussian window filter was applied to smoothly suppress the field $|E_z|$ at the boundaries of the selected region. This approach minimizes edge artefacts and reduces spectral leakage in momentum space. Following this step, a two-dimensional FFT of the complex near-field distribution was calculated (Figure S2b). In reciprocal space, distinct momentum components corresponding to the propagating hyperbolic phonon polaritons become visible as high intensity regions at wavevectors determined by the polariton dispersion.

To isolate the contribution of specific polaritonic modes, circular band-pass filters were applied around the relevant momentum-space features to suppress experimental noise as well as slowly varying background offsets originating from the measurement process. These same processing steps were applied to simulations in order to keep a consistent method. The filtered reciprocal-space data were then transformed back into real space using the inverse FFT to recover the spatially resolved complex near-field distribution associated exclusively with the selected polaritonic modes. From this reconstructed complex field, both $|E_z|$ and $\varphi_z$ were extracted for further analysis (Figure S2c). From Maxwell's equations using boundary conditions (see Supplementary Note 3), the in-plane field of the topological texture (Figure S2d) can be reconstructed in a straightforward manner. Finally, topological features were isolated spatially to calculate the corresponding skyrmion number density SND (Figure S2e) and topological invariants (Figure S2f).

## Supplementary Note 3: Relationship between the out-of-plane electric field and the corresponding in-plane field distribution

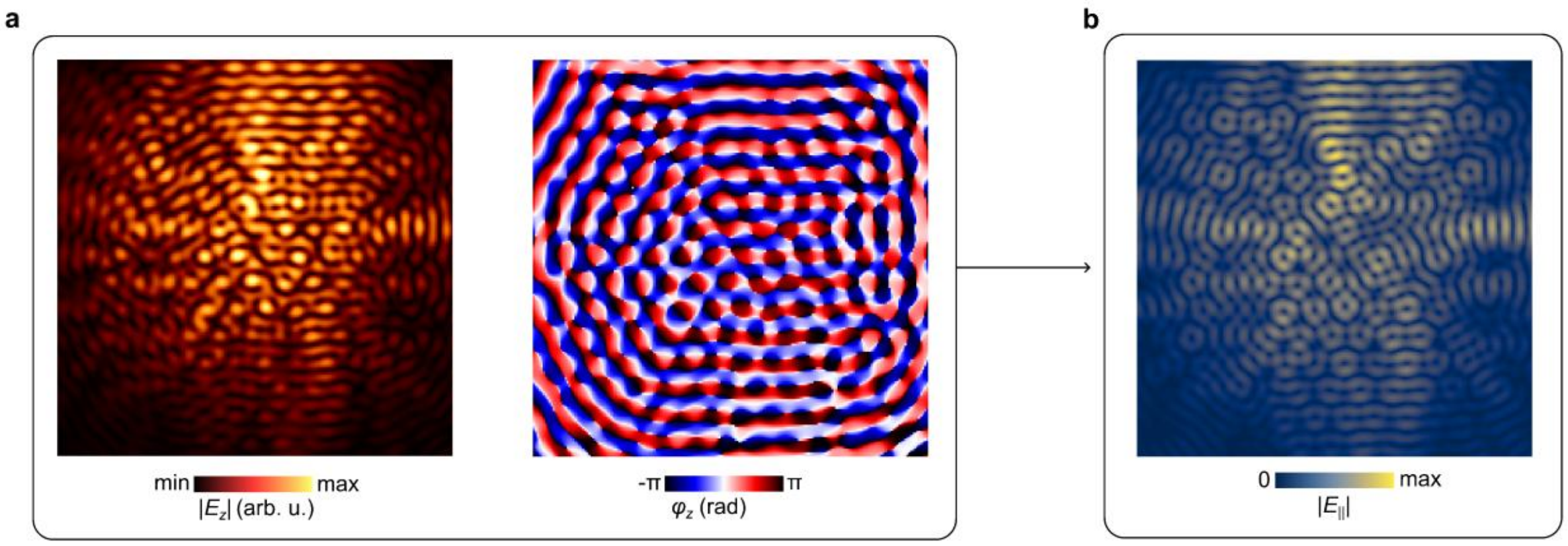

***Figure S3: Relationship between the out-of-plane electric field and the corresponding in-plane field distribution.*** *a) Experimental out-of-plane electric field amplitude $|E_z|$ (left) and phase φ (right) from the measurement shown in Figure 3c. b) Calculated magnitude of the corresponding in-plane electric field, $|E_{/\!/}|$ obtained from inserting measured Ez into Equation E1.*

Neglecting intrinsic material losses and assuming harmonic time dependence, structured electromagnetic near fields can be described as a coherent superposition of multiple evanescent waves. Considering N waves of equal amplitude $E_{z,0}$, the out-of-plane electric field is expressed as:

$$E_z = \sum_{n=1}^{N} E_{z,0} e^{ik_{||,n}\cdot r_{||} + \varphi_{||,n}} e^{-\kappa_n \cdot z} \quad \text{(E1)}$$

with $r_{||} = (x, y)$ and $\kappa_n = \sqrt{k_{||,n}^2 - k_0^2}$. The in-plane wavevectors $k_{||,n}$determine the propagation directions of the individual evanescent waves, while their relative phases determine the symmetry and spatial arrangement of the resulting interference pattern. Both the wavevectors and their phases can be simultaneously extracted directly from spatial maps resulting from s-SNOM measurements in pseudo-heterodyne detection scheme.

Furthermore, for TM-polarized evanescent waves, the in-plane electric field components are not independent quantities but are directly coupled to the out-of-plane field through Maxwell's equations and the transversality condition. Consequently, the in-plane fields can be obtained directly from the spatial derivatives of the out-of-plane components[4]:

$$\begin{pmatrix} E_x \\ E_y \end{pmatrix} = \frac{1}{k_{||}^2} \begin{pmatrix} \frac{\partial^2 E_z}{\partial x \partial z} \\ \frac{\partial^2 E_z}{\partial y \partial z} \end{pmatrix} = \frac{-\kappa_n}{k_{||}^2} \begin{pmatrix} \frac{\partial E_z}{\partial x} \\ \frac{\partial E_z}{\partial y} \end{pmatrix} \tag{E2}$$

This relationship demonstrates that every spatial variation in the out-of-plane field is accompanied by a corresponding rotation of the in-plane electric field. As several evanescent waves interfere, the resulting vector field evolves smoothly through space, allowing the local field orientation to wrap around singular points and produce non-trivial topological textures, such as skyrmions and merons.

The symmetry of the excitation determines the topology of the resulting field. Six-fold symmetric excitation produces hexagonal polaritonic skyrmion lattices, whereas four-fold symmetry generates square meron lattices. As the characteristic length scale of these interference patterns is governed by the in-plane wavevector $k_{||}$, which can be up to 50-100 times larger than the wavevector of the incident light for bulk hyperbolic polaritons in hBN[2], the resulting topological field distributions are produced with deeply subwavelength dimensions.

## Supplementary Note 4: Stability of experimentally measured skyrmion under global phase evolution

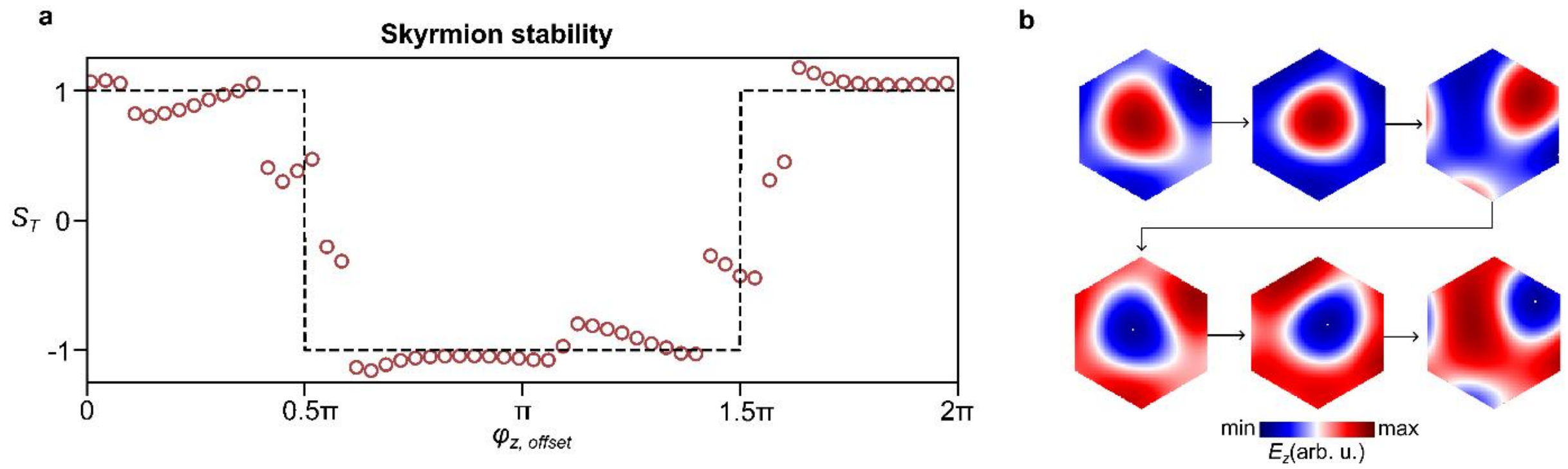

***Figure S4. Stability of experimentally measured skyrmion under global phase evolution.*** *a) Evolution of the calculated skyrmion number $S_T$ for the experimentally measured skyrmion shown in Figure 3a as a uniform phase offset $\varphi_{z,offset}$ is applied to the out-of-plane electric field. The topological invariant remains quantized over extended phase intervals, confirming the robustness of the skyrmion against global phase perturbations. b) Selected snapshots of the corresponding out-of-plane electric field for different applied phase offsets, demonstrating the continuous deformation of the field texture across the phase variation.*

## Supplementary Note 5: Comparative study of s-SNOM for characterization of polaritonic field topological lattices

| Author | Coupling structure | | Detection scheme | | Tuna-bility | Confine-ment |
|---|---|---|---|---|---|---|
| | Structure | Side length illumination and polariton wavelength | Polari-zation of illu-mination | Detection | | |
| Our work | Optically programmed hexagon and rectangle in IST<br>**Wavelength independent** | a~3-1μm<br>$\lambda_0$=6.8-6.7μm<br>$\lambda_p$=239-168nm | Linear | Reflection SNOM | yes | $\lambda_0$/40 |
| Tsesses et al. 2018[4] | 6 grating slits milled into gold film, arranged in hexagon, one side offset by $\lambda_p$/2<br>**Wavelength dependent** | a~5μm;<br>$\lambda_0$=660nm<br>$\lambda_p$=636nm | Circular | Trans-mission SNOM | no | $\lambda_0$/1.04 |
| Mangold et al. 2026[5] | 6 chromium ridges arranger in hexagon on SiC membrane, each offset akin to archimedean spiral $\lambda_p$/6<br>**Wavelength dependent** | a~45-15μm<br>$\lambda_0$ =12- 11μm<br>$\lambda_p$=12-3μm | circular | Trans-mission SNOM | no | $\lambda_0$/3.6 |
| Baú et al. 2026[6] | Metasurface consisting of arrays of dielectric hexagonal resonators hosting broad resonance due to quasi bound states in the continuum<br>**Wavelength independent** | $\lambda_0$ =6.6-6.5μm<br>$\lambda_p$=451-271nm | linear | Trans-mission SNOM | yes | $\lambda_0$/25 |

***Table S1. Comparative study of s-SNOM for characterization of polaritonic field topological lattices.***

A comparative study on previous experimental observation of polaritonic field topological lattices utilizing s-SNOM is summarized in Table S1. Note, further realization of polaritonic topological lattices based on spin fields exist but are not reviewed here. We highlight the differences in coupling structure and detection scheme.

### Comparison on previously utilized coupling structures and detection schemes

Tsesses et al.[4] utilized six grating slits milled into a gold film and arranged in a hexagon to launch surface plasmon polaritons in gold. One of the grating slits is offset by half of the polariton wavelength to compensate for the phase difference due to circularly polarized illumination. Consequently, this coupling structure is wavelength dependent. The coupling structure is much larger than the illumination wavelength (a ≈ 5 μm >> $\lambda_0$ ≈ 660 nm). The polaritons in gold are weakly confined and their wavelength is close to that of the illumination wavelength ($\lambda_0$ ≈ 660 nm ~ $\lambda_p$ ≈ 636 nm). The polaritonic interference pattern is detected with s-SNOM in transmission mode.

Mangold et al.[5] utilized six chromium ridges arranged in a hexagon to launch surface phonon polaritons in a SiC membrane. Each of the ridges is offset by a sixth of the polariton wavelength akin to an Archimidean spiral to compensate for the phase difference due to circularly polarized illumination. The coupling structure is larger than the illumination wavelength (a ≈ 15-45 μm > $\lambda_0$ ≈ 11-12 μm). The confinement of the polaritons in the SiC membrane depends on the illumination wavelength and can be tuned from weak ($\lambda_0$ ≈ 12 μm ~ $\lambda_p$ ≈ 12 μm) to high ($\lambda_0$ ≈ 11 μm > $\lambda_p$ ≈ 3 μm) confinement. The polaritonic interference pattern is detected with s-SNOM in transmission mode.

Baú et al.[6] utilized a completely different coupling scheme not relying on the launching polaritons as explained in the main text. Instead, they utilized a metasurface supporting quasi-bound states in the continuum resonances with homogeneous fields across the single resonator structures of the metasurface that generate the polaritonic topological lattices. By tuning the frequency within the quasi-bound states in the continuum resonance, tuning the size of the topological lattice site has been demonstrated. The resonance of the metasurface is excited with linearly polarized light. As in our work highly confined hyperbolic phonon polaritons in hBN ($\lambda_0$ ≈ 6.5-6.6 μm ≫ $\lambda_p$ ≈ 271-451 nm) are used for the polaritonic topological lattice. The polaritonic interference pattern is detected with s-SNOM in transmission mode.

In our work, we utilize coupling structures optically programmed into IST to launch hyperbolic phonon polaritons in hBN overcoming the need for cumbersome fabrication schemes such as ion beam milling and lithography needed to create the previously employed coupling structures. Further, we utilize coupling structures illuminated with linearly polarized light that do not compensate for a phase shift. Hence, our structures are wavelength independent and allow for tuning the illumination wavelength and thereby for tuning the confinement without adaption of the coupling structure. One plausible explanation might be that our coupling structures are smaller than the illumination wavelength ($a \approx$ 1-3 µm $< \lambda_0 \approx$ 6.7-6.8 µm) contrasting the previous approaches that utilized coupling structures larger than the illumination wavelength ($a > \lambda_0$). The hyperbolic phonon polaritons in hBN are highly confined allowing for a confinement of $\lambda_0/40$. We detect the polaritonic interference pattern with s-SNOM in reflection mode whereas the previous realizations utilized s-SNOM in transmission mode.